\documentclass[11pt]{article}
\pdfoutput=1
\usepackage{graphicx,rotating,slashed,amsmath,xcolor,amssymb,amsfonts,colortbl,cite,bbold}
\usepackage{jheppub}
\usepackage{mathrsfs}
\usepackage{xcolor}
\usepackage{caption}
\usepackage{subcaption}
\usepackage{physics}

\def\gsim{ \lower .75ex \hbox{$\sim$} \llap{\raise .27ex \hbox{$>$}} }
\def\lsim{ \lower .75ex \hbox{$\sim$} \llap{\raise .27ex \hbox{$<$}} }
\def\be{\begin{equation}}
\def\ee{\end{equation}}
\def\bea{\begin{eqnarray}}
\def\eea{\end{eqnarray}}

\newcommand{\ba}{\begin{array}}
	\newcommand{\ea}{\end{array}}

\newcommand{\commentout}[1]{}

\newcommand{\comment}[1]{}

\newcommand{\bs}{\begin{split}}

	\def\ba{\begin{eqnarray}}
	\def\ea{\end{eqnarray}}

	\def\({\left(}
	\def\){\right)}

	\definecolor{jn}{RGB}{10, 10, 200} 
	\definecolor{js}{RGB}{204, 0, 0} 
	\definecolor{pgf}{RGB}{10, 150, 10} 

	\newlength{\stheight}
	\newcommand\textst[1][fu-grey]{
		\ifmmode\setlength{\stheight}{+1.0ex}
		\else\setlength{\stheight}{+0.5ex}
		\fi
		\bgroup\markoverwith{\textcolor{#1}{\rule[\the\stheight]{2pt}{1.0pt}}}\ULon
	} 
	\newcommand{\textins}[2][fu-grey]{
		\ifmmode\mathcolor{#1}{#2}
		\else\textcolor{#1}{#2}\@\,
		\fi
	}
	\graphicspath{{./}}
	\allowdisplaybreaks
	
\providecommand{\abs}[1]{\lvert#1\rvert} 
\newcommand{\vect}[1]{\mathbf{#1}} 
\newcommand{\pd}[1]{\partial#1} 

\begin{document}

	\title{Effective  Cuscuton Theory}
	
 	\author[a,b]{Maria Mylova}
 	\author[c,d,e]{, Niayesh Afshordi}

    \affiliation[a] {Kavli Institute for the Physics and Mathematics of the Universe (WPI), The University of Tokyo, Kashiwa, Chiba 277-8583, Japan}
    \affiliation[b] {Science Education, Ewha Womans University, 52 Ewhayeodae-gil, Seoul, Republic of Korea}

 	\affiliation[c]{Department of Physics and Astronomy, University of Waterloo, Waterloo, ON, N2L 3G1, Canada}
 	\affiliation[d]{Waterloo Centre for Astrophysics, University of Waterloo, Waterloo, ON, N2L 3G1, Canada}
 	\affiliation[e]{Perimeter Institute for Theoretical Physics, 31 Caroline St. N., Waterloo, ON, N2L 2Y5, Canada}

\emailAdd{mylova@g.ecc.u-tokyo.ac.jp}
\emailAdd{nafshordi@pitp.ca}

	\date{today}
	 \abstract{Cuscuton field theory is an extension of general relativity that does not introduce additional propagating degrees of freedom, or violate relativistic causality. We construct a general geometric description of the cuscuton field theory by introducing curvature corrections to both the volume (potential) and the surface (kinetic) terms in the original cuscuton action. Our assumptions involve a stack of spacelike branes, separated by 4-dimensional bulks. We conjecture that the cuscuton, initially a discrete field, becomes continuous in the limit, there are many such transitions. From this we derive an effective action for the cuscuton theory and show that at the quadratic level our theory propagates only the two tensorial degrees of freedom. 
	}
	
	\maketitle
	

	\section{Introduction}

 Lovelock's theorem \cite{Lovelock:1971yv, Lovelock:1972vz} states that within the confines of four-dimensional spacetime, General Relativity (GR) is the sole metric theory that propagates two tensorial degrees of freedom, while at the same time satisfies full diffeomorphism invariance and produces field equations with at most two derivatives acting on the metric. However, due to the dynamical cosmological background, our existent universe spontaneously breaks time-diffeomorphism invariance.
 In light of this, it is natural to ask whether the physics 
 that aim to elucidate our present-day cosmological mysteries could be better described by a theory which modifies GR but, nevertheless,   propagates only two tensorial degrees of freedom? 

This line of inquiry has received a lot of attention over the past decade, 
originating with the seminal cuscuton theory
\cite{Afshordi:2006ad}. Since its inception, there have been several attempts to formulate cuscuton-like theories, which go under the names of extended cuscuton theory \cite{Iyonaga:2018vnu}, minimally modified gravity (MMG) \cite{DeFelice:2015hla, Lin:2017oow, Mukohyama:2019unx,Aoki:2018brq, DeFelice:2020eju, Aoki:2020lig, Aoki:2021zuy,DeFelice:2022uxv} and spatial covariant gravity with two tensorial degrees of freedom (TTDOF) \cite{Gao:2019twq, Hu:2021yaq}, to mention a few.  The key motivation behind this ``minimalist" approach to modified gravity is to reconcile disparities between theory and empirical observations in the so-called dark sector without resorting to the introduction of extra degrees of freedom.
 There have been a number of studies on cuscuton-like theories revealing intriguing theoretical and observational ramifications for both cosmology and astrophysics.
A detailed review is beyond the scope of this work, but the curious reader can explore \cite{Afshordi:2007yx, Magueijo:2008pm, Afshordi:2009tt, Bessada:2009ns, Afshordi:2014qaa, Agarwal:2014ona, DeFelice:2015moy, Bhattacharyya:2016mah, Afshordi:2016guo, Boruah:2017tvg, Chagoya:2018yna, Boruah:2018pvq, DeFelice:2018vza, Aoki:2020oqc, Quintin:2019orx, DeFelice:2021trp, Aoki:2020iwm, Aoki:2020ila, Iyonaga:2020bmm, DeFelice:2020onz, DeFelice:2020prd, DeFelice:2020cpt, Lin:2020uvx, Panpanich:2021lsd, Mylova:2021eld, deAraujo:2021cnd, Pookkillath:2021gdp, Bartolo:2021wpt, Iyonaga:2021yfv, Maeda:2022ozc, Hiramatsu:2022ahs, Kohri:2022vst, Saito:2023bhn, Yao:2023qjd} and the references therein.

Much progress has been made in the classification of cuscuton-like theories, whose derivation usually involves a level of brute force while at the same time obeying some organising principle. 
The standard prescription is to assume a generic Lagrangian that obeys invariance under spatial diffeomorphism. It is a general function of the lapse $N$, the intrinsic curvatures $\mathcal{R}_{ij}$, and the extrinsic curvatures $K_{ij}$. The Lagrangian may also contain functions of time $t$ and/or spatial derivatives $D_i$. Then one can try to derive consistency conditions motivated by a Hamiltonian analysis \cite{Lin:2017oow, Gao:2019twq} that ensure the only dynamics in the theory are due to the two tensorial degrees of freedom. 
Another approach is instead to start directly from the Hamiltonian of the theory and determine which conditions need to be satisfied so that the scalar degree of freedom can be eliminated. Then one can build a Lagrangian which is made up of terms which satisfy these conditions \cite{Mukohyama:2019unx, Yao:2020tur, Yao:2023qjd}. Alternatively, perturbative methods can be used to impose conditions that no scalar degree of freedom propagates at each order in the perturbative expansion \cite{Hu:2021yaq}. All of these approaches have their merits and shortcomings. Conditions imposed at the level of the Hamiltonian can be very useful due to their generality but at the same time challenging to solve in order to produce viable Lagrangians. Meanwhile, the simple structure one obtains when starting directly with a Hamiltonian analysis may be difficult to translate into a concrete Lagrangian. In contrast, perturbative methods can be more directly applied to the Lagrangian but may involve a certain level of tuning, albeit finite.

In this work, we adopt an alternative approach. We are interested in constructing an effective action for the cuscuton by taking an effective field theory (EFT)  point of view. This requires us to extend the cuscuton leading action, order by order in a perturbative gradient expansion. One difficulty with doing so is that one is required to construct an effective field theory for a non-dynamical degree of freedom. Indeed, the typical formulation of effective field theories relies on having knowledge of the fundamental symmetries and available degrees of freedom. Conversely, the cuscuton lacks internal dynamics. However, what about its underlying symmetries? In flat spacetime, the cuscuton exhibits a scalarless symmetry \cite{Chagoya:2016inc, Tasinato:2020fni}, guaranteeing the absence of propagating scalar excitations. Viewing it through a braneworld perspective, its action can be derived from a brane probing a 5-dimensional bulk with two time-like directions. This hints at the possibility of expressing the cuscuton as a geometric theory of defects. Nevertheless, so far, no one has succeeded in covariantizing this symmetry.

The only remaining avenue is to extend the cuscuton theory based on its geometric description, originally presented in \cite{Afshordi:2006ad}.  This can become manifest by re-expressing the cuscuton action as a combination of surface and volume contributions. For a scalar field  with a time-like gradient $\partial_\mu\phi$, the cuscuton action (\ref{eq:cusc1}) can be written as\footnote{For more details see \cite{Afshordi:2006ad} which is where the geometric picture of the cuscuton originated.}
\begin{equation}\begin{split}
S_\phi  = \int \dd^4{x} \sqrt{-g} \qty[\mu^2 \abs{n^\mu \pd_\mu \phi} - V(\phi)]
&= \mu^2 \int \dd{\phi} \Sigma(\phi) - \int \dd^4{x} \sqrt{-g} V(\phi),
\\ \qq{with} \Sigma(\phi) &\equiv \int_\phi d^3x \sqrt{\gamma},
\label{eq:cuscgem}
\end{split}\end{equation}
where $\gamma$ is the determinant of the induced metric on the hypersurface of constant $\phi$.
This way, it becomes apparent that the volume and surface   terms represent the potential and the leading order kinetic term of the cuscuton, respectively. 
Thus, from an effective field theory point of view, one conceivable approach would be to try and extend (\ref{eq:cuscgem}) by introducing curvature corrections to both volume and surface terms. 

The action in Eq. (\ref{eq:cuscgem}) can be understood as analogous to the surface and volume terms, associated with the energy of soap bubbles and films in a 3-d Euclidean space. For cuscuton, the 2-d surfaces of soap films are replaced by 3-dimensional spatial hypersurfaces of constant $\phi$, living in a 4-dimensional Lorentzian spacetime (or bulk). These spacelike branes or S-branes are understood to be topological defects, localised on a spatial hypersurface representing an instant in time (in a particular coordinate system). They are usually taken to have no thickness but only a surface. 
In general, S-branes have been studied in relation to string theory \cite{Gutperle:2002ai, Cicoli:2023opf},  but we do not make this connection here. Taking as our starting point the geometric description in (\ref{eq:cuscgem}), in the new geometric description, corrections to the cuscuton kinetic term reside on the boundaries or S-branes of the 4-dimensional picture, while the corrections to the potential live in the $4$-dimensional bulk. This setup is reminiscent to braneworld constructions \cite{deRham:2010eu, Goon:2011qf} which are typically used to build actions with higher derivative terms which admit at most quadratic equations of motion and therefore do not propagate any extra degrees of freedom around the background. In braneworld scenarios, a dynamical 3-brane is moving in a fixed 4+1 dimensional background. Our setup is different in that:

\medskip

$\bullet$ We assume a stack of spacelike branes living in a 4-dimensional spacetime. 
These are discrete spatial surfaces that do not talk to each other (i.e., if we move one, the other does not know). This results in the non-dynamical nature of the cuscuton field.  

\medskip

$\bullet$ We further assume that the cuscuton is a discrete field (labelling each transition) that becomes continuous in the limit that there are many such transition interfaces, assuming the geometry varies slowly from one transition to the next.

\medskip

$\bullet$ To realise several such transitions, we expect the branes to interface between different vacua. This implies that the bulks separated by the S-branes may have different couplings, e.g., different cosmological and gravitational constants. 

\medskip

In this investigation, we exclusively concentrate on Lovelock terms, which include the cosmological constant, the Einstein-Hilbert action, and higher-order Lovelock terms like the Gauss-Bonnet term in 4 dimensions. Interestingly, while the Gauss-Bonnet term is topological in 4 dimensions, we observe that in the continuous limit of these discrete transitions, this term plays a role in the overall dynamics. The Lovelock terms existing within the bulk have associated boundary terms located on the surfaces of the branes.  In the continuous limit, we discover that these terms function as ``counterterms'', ensuring that the metric carries only two degrees of freedom.

To summarise the above, in this discrete geometric picture, the cuscuton is represented by the mismatch or discontinuities in the couplings of Lovelock terms, as we jump from one bulk to another. The cuscuton effective action emerges in the continuous limit of these discrete transitions. This process is explained in detail in section \ref{ECT} and draws parallels with early attempts to formulate effective actions for the study of critical phenomena in condensed matter systems \cite{Wilson:1971bg}. In this sense, our effective construction may initially appear to be crude and lacking finesse. Nevertheless, it could be useful to demystify the puzzles surrounding the cuscuton and cuscuton-like theories in general. At this stage, it is best to consider our setup as a pure mathematical tool for building an effective cuscuton theory. We comment more on this in section \ref{discus}.

Before closing this section, it is worth mentioning that non-propagating fields appear in other places in the literature \footnote{We thank C. P. Burgess for bringing these topics to our attention.}. Non-propagating form fields are  known to play an important role in the EFT for Quantum Hall (QH) systems, where in this case non-propagating
Maxwell fields (2-forms) are essential in expressing how the QH
state samples the topology of the space in which the system resides \cite{burgess2020introduction}.  Another example comes from the effective construction in  Supersymmetric Large Extra Dimensions (SLED) models, where a non-propagating 4-form is what tells the EFT that
flux in the extra dimensions is quantized \cite{Burgess:2015lda}. The presence of this field imposes a particular form on the low-energy potential that is precisely what is
required for it to reproduce the dimensionally reduced result. Finally, in \cite{Bielleman:2015ina} they show that all auxiliary fields in
the low energy 4-dimensional EFT of string compactifications arise as 4-forms. This implies there is a very general
connection in 4 dimensions between non-propagating topological form fields and
non-propagating auxiliary fields in 4-dimensional supergravity. While these topics are not often discussed, they may be especially important for the technical naturalness of theories (see \cite{Burgess:2021juk, Burgess:2021obw} for possible applications).  Although these subjects are interesting to explore, they come with a plethora of technical intricacies associated, e.g., with supersymmetry and higher dimensions. On the other hand,  the cuscuton serves as a straightforward illustration of a 4-dimensional non-propagating field, offering a more accessible avenue towards understanding how non-propagating fields may lose their dynamics.

This manuscript is organised as follows: first, we give a brief introduction of the cuscuton in Section \ref{cuscintro} and  lay down the main ingredients for the construciton of the effective cuscuton theory in section \ref{ECT}. We proceed to analyse the leading terms of the effective cuscuton action in Section \ref{ADM}. We briefly discuss the treatment of higher-order operators in Section \ref{beyond} and we conclude in Section \ref{discus}.

\section{A brief introduction of the Cuscuton}
\label{cuscintro}

Historically, cuscuton was introduced as a scalar field characterised by an infinite speed of sound $c_s \to \infty$. However, it does not propagate information outside (or inside) the light cone because it lacks a dynamical phase space. From a quantum perspective, the energy of a cuscuton quantum excitation is given by $\hbar\omega= \hbar c_s k \to \infty$, for any finite wavenumber $k$, which implies that it takes infinite energy to carry information using cuscuton excitations.   Furthermore, it benefits from protection against radiative corrections thanks to an enlarged set of symmetries \cite{Pajer:2018egx, Grall:2019qof}. 

However, we should note that the limit $c_s \to \infty$ is only well defined for a restricted set of boundary conditions. From a geometric perspective, as we see below, the cuscuton constant field surfaces are constant mean curvature surfaces (or ``Lorentzian soap bubbles'') that are anchored on the boundary; healthy boundary conditions would require that the surfaces do not intersect in the bulk \cite{Afshordi:2006ad}. Furthermore, an initial-boundary value problem for such a system is only well-posed in the frame that the cuscuton is uniform, which ensures elliptic constraint equations (rather than pathologies such as apparent instabilities, or instantaneous modes that propagate back in time \cite{DeFelice:2018ewo}).

The  action is expressed in terms of a scalar field $\phi$ with a non-canonical kinetic term, as 
\begin{equation}\begin{split}
S = \int\dd^4{x} \sqrt{-g} \qty(\frac{M_P^2}{2} R - \mu^2 \sqrt{ X} + V(\phi) ), \quad X= - g^{\mu\nu}\pd_\mu \phi \pd_\nu \phi, \quad X > 0.
\label{eq:cusc1}
\end{split}\end{equation}
The cuscuton is a non-dynamical field satisfying constraint equations that modify the dynamics of the fields it couples to. To see this, let us assume a flat Friedmann-Lemaitre-Robertson-Walker (FLRW) spacetime with metric
\begin{equation}\begin{split}
\dd s^2 = - \dd t^2 + a^2(t) \, \dd x^2,
\label{eq:metric}
\end{split}\end{equation}
and take a look at the background equations of motion, which read
\begin{align}
\label{fried1}
   & H^2 = \frac{1}{3} V(\phi),
    \\ 
    \label{fried2}
   & \dot{H} = -\frac{3 H^2}{2} - \frac{1}{2} \mu^2 \sqrt{\dot\phi^2} + \frac{1}{2} V(\phi),
   \\ 
   & 3 \mu^2 H = - V'(\phi),
\end{align}
where the Friedmann equations  are given by (\ref{fried1}) and (\ref{fried2}), while  the last expression comes from varying the action with respect to the scalar field. It is now clear that the scalar field equation can be represented by a straightforward algebraic expression.
We can then solve for the first Friedmann equation to get $H = \sqrt{V(\phi)/6}$ and substituting into the last equation, we find $\mu^2 \sqrt{3 V(\phi)} = - V'(\phi)$. Therefore, we see that the equations of motion for GR+cuscuton cannot be generally satisfied, and thus we need to add additional content, i.e. matter, to the theory 
\begin{equation}\begin{split}
H^2 = \frac{\rho_{tot}}{3},
\label{eq:cusc2}
\end{split}\end{equation}
where $\rho_{tot} = \rho_m + V(\phi)$ with $M_P = 1$, to find instead that
\begin{equation}\begin{split}
\frac{1}{3\mu^4} V'^2(\phi) - V(\phi) = \rho_m,
\label{eq:cusc3}
\end{split}\end{equation}
where $\rho_m$  is the background density of matter species \cite{Afshordi:2007yx}. It is now clear that the cuscuton has no internal dynamics, but it merely  modifies the fields it couples to through constraint equations. As an example, if we fix $\rho_m$, this consequently fixes $\phi$, from which it follows that $\phi$ is just a function of matter density. 

The cuscuton has an interesting geometric representation which was first pointed out  in \cite{Afshordi:2006ad}. Its equation of motion can be written in a covariant form, as
\begin{equation}\begin{split}
\frac{1}{\sqrt{-g}} \pd_\mu\qty[\sqrt{-g} \frac{g^{\mu\nu} \pd_\nu \phi}{\sqrt{X}}] + V'(\phi) =0.
\label{eq:cuscgem1}
\end{split}\end{equation}
Using that the normal vector $n^\mu$ for constant $\phi$ hypersurfaces, is given by 
\begin{equation}\begin{split}
n^\mu \equiv \frac{\pd^\mu \phi}{\sqrt{X}},
\end{split}\end{equation}
and the definition of the extrinsic curvature $K = \nabla_\mu n^\mu$, we find that (\ref{eq:cuscgem1}) becomes 
\begin{equation}\begin{split}
K(\phi) = -\frac{V'(\phi)}{\mu^2}.
\label{CMC1}
\end{split}\end{equation}
Therefore, the constant $\phi$ hypersurfaces are also surfaces of constant mean curvature (or CMCs).  The geometric picture that arises through this set-up is akin to soap bubbles or films in 3-dimensional  Euclidean space. In this case, the CMC condition can be seen as the balance between the pressure and the surface tension forces. This can be understood by rewriting (\ref{CMC1}) in terms of the pressure difference across the surface, as
\begin{equation}\begin{split}
(\mu^2\Delta \phi) K_{\text{disc}} = -\Delta V.
\label{CMC2}
\end{split}\end{equation}
We thus see that $\mu^2\Delta \phi$ can be interpreted as the surface tension of the soap bubble. 
Eq. (\ref{CMC2}) can be solved to find the surface of the soap bubble or film. The result is a discrete set of solutions which can be determined by supplying the necessary boundary conditions.

From this we understand that the  contribution to the kinetic term in (\ref{eq:cuscgem}) can be thought of as summing over all such surfaces and taking the continuous limit. This motivates a natural conjecture that the cuscuton is a fundamentally discrete field associated with spacelike discontinuities/ topological defects in spacetime.

\section{Effective cuscuton  theory }
\label{ECT}
	
	\subsection{Constructing the effective action}

 As we explained in the Introduction, the aim of this work is to extend the geometric picture above into an effective cuscuton theory, that generalises the surface and volume terms. We do so by putting forward the following conjecture: 

 \medskip

 \textit{The cuscuton is a fundamentally discrete `field' labelling transitions in spacetime. In the limit of many transitions, it can be approximated by a continuous field with an infinite speed of sound. In this way, the effective cuscuton theory (ECT) emerges as a curvature expansion of the geometric theory of a series of bulks and their boundaries.
}

 \medskip

We now explain how this picture can be realised. We consider a stack of spatial branes, having zero thickness, labelled $\phi=1,2,\cdots$. These can be thought of as hypersurfaces, each representing an \textit{instant in time}.  For the sake of simplicity, we take only gravity to propagate in the bulk (i.e., we turn the standard-model fields off). Assuming that these branes live on some bulk geometry, we expect the kinetic term to be given by the surface, whereas the potential will be given by the volume in the geometric picture. Additionally, we expect that geometric terms due to 4-d bulk gravity, i.e., the Einstein-Hilbert term and higher-order curvature invariants (for example, Lovelock invariants, such as the non-dynamical Gauss-Bonnet in 4 dimensions), will have associated boundary terms living on the brane geometry (see \cite{deRham:2010eu, Goon:2011qf, Dyer:2008hb} and references therein).

\begin{figure}[htbp]
		\begin{center}			\includegraphics[width=0.80 \textwidth]{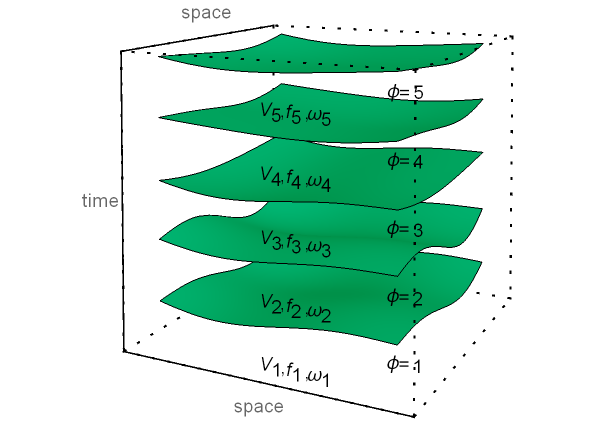}
			\caption{\it The geometric picture of the cuscuton is described by a stack of non-identical spatial branes labeled by $\phi =1 \cdots n$. Each bulk has its own value for the cosmological constant $V_\phi$, its own gravitational strength $f_\phi$, and its own value for the Lovelock couplings $\omega_\phi$.}
		\label{fig:1}
		\end{center}
	\setlength{\abovecaptionskip}{0pt plus 1pt minus 1pt}
	\end{figure}

 An interpretation of this geometrical description, which can be seen in Fig.(\ref{fig:1}), is that we are essentially studying the theory for consequent $3$-dimensional defects in a $3+1$ dimensional field theory, with the key idea being,  the cuscuton describes discontinuous jumps in spacetime.  In this picture, we focus on the continuous limit of these discrete transitions (i.e., we approximate the sum as an integral over the geometric invariants living on the bulks and branes).
As long as constant-$\phi$ hypersurfaces are spacelike, we can choose the unitary gauge, such that the scalar field is homogeneous, $\phi=\phi(t)$, and thus its gradient points in the time direction. 

We first give a prescription for how to realise the continuous limit of this discrete geometric picture. 
The discrete action is given by the sum over all the geometric terms living in the bulks and the branes. For simplicity, let us focus only on the volume and surface contributions to the discrete action, which can be written as
\begin{equation}\begin{split}
S_{\text{disc}} = \sum_\phi  {\cal V}(\phi,\phi+\Delta\phi)\times V(\phi) + \sum_\phi   {\cal S}(\phi)\times  c_1(\phi)  \Delta \phi,
\label{disc}
\end{split}\end{equation}
where ${\cal V}(\phi,\phi+\Delta\phi)$ is the volume enclosed between $\phi$ and $\phi+\Delta\phi$ surfaces, while ${\cal S}(\phi)$ is the area of the $\phi$-hypersurface.
Taking the continuous limit for the volume term gives
\begin{equation}\begin{split}
 \sum_\phi  {\cal V}(\phi,\phi+\Delta\phi)\times V(\phi)  \Big|_{\Delta\phi \rightarrow 0} = \int \dd^4{x} \sqrt{-g} ~V(\phi), 
\label{disc1}
\end{split}\end{equation}
assuming that  $V(\phi)$ is a slow-varying function of $\phi$. This is nothing more than the potential $V$ for the cuscuton. Similarly, taking the continuous limit of the surface term gives
\begin{equation}\begin{split}
 \sum_\phi   {\cal S}(\phi)\times  c_1(\phi)  \Delta \phi \Big|_{\Delta \phi \rightarrow 0} = \int  \dd^3 x \, \dd\phi \, \sqrt{\gamma}   \, c_1(\phi),
\label{disc2}
\end{split}\end{equation}
where $\sqrt{\gamma} \, \dd^3{x}$  is the  3-dimensional invariant surface element and $c_1(\phi)$ is a slow-varying function of $\phi$.
To recover the 4-dimensional volume element we need to simply multiply and divide by $N \dd t$, where $N$ is the lapse. Therefore, the surface contribution to the action becomes 
\begin{equation}\begin{split}
 \int  \dd^3 x \, \dd\phi  \,  \sqrt{\gamma} \,  c_1(\phi)= \int  \dd^4 x \, \sqrt{-g} \, c_1(\phi)  \frac{1}{N} = \int  \dd^4 x \, \sqrt{-g} \, c_1(\phi)  \sqrt{X}.
\label{disc3}
\end{split}\end{equation}
This way, we recover the cuscuton kinetic term\footnote{This can be understood because in the unitary gauge the role of the physical clock is played by $\phi$ such that $\nabla_\mu$ is time-like. Therefore, $X = - g^{\mu\nu} \pd_\mu t \pd_\nu t = 1/N^2$.} from the surface contribution in the geometric picture\footnote{Comparing to Eq. (\ref{eq:cusc1}), we see that $\mu^2= c_1(\phi)$ matches the original cuscuton action.}. 
From this it follows that all other possible gravitational terms living in the boundary will, naturally, be coupled to the cuscuton through $\sqrt{X}$.  

Following the same prescription, we can now extend the theory by summing over the geometric invariants living in the bulks and on the boundaries, ordered in powers of curvature. In the continuous limit $\Delta\phi \to 0$, the sum becomes an integral, which we call the effective cuscuton action (ECT):  
\begin{equation}\begin{split}
S_{{ECT}} & =  \Lambda^4\int \dd^4x \sqrt{-g}  \bigg[ V(\phi)  + f(\phi) \frac{R}{2 \Lambda^2} + \omega(\phi) \frac{R_{GB}}{2 \Lambda^4}
\\& + c_1(\phi) \frac{\sqrt{X}}{\Lambda} +  \varepsilon c_2(\phi)  \frac{\sqrt{X} K}{\Lambda^2} + \frac{c_3(\phi)  \sqrt{X} \, \mathcal{R}}{\Lambda^3} +  c_4(\phi)  \frac{\sqrt{X} \,  \mathcal{K}_{GB}}{\Lambda^4} + \cdots\bigg],
\label{eq:cw2a}
\end{split}\end{equation}
where, from now on, we assume $\phi$ is dimensionless, and introduce a mass scale $\Lambda$, associated with the UV-completion of cuscuton, which could be comparable or lower than the Planck mass. Here, ``$+\cdots$'' indicates higher-order curvature invariants that may live in the bulk and/or on the boundary. $R$ is the $4$-dimensional Ricci scalar and $R_{GB}$ is the $4$-dimensional Gauss-Bonnet term, given in terms of the $4$-dimensional curvature invariants:
\begin{equation}\begin{split}
R_{GB} = R^2 - 4 R^{\mu\nu} R_{\mu\nu} + R^{\mu\nu\rho\sigma} R_{\mu\nu\rho\sigma}.
\label{eq:GB}
\end{split}\end{equation}
The potential  $V(\phi)$ of the scalar field corresponds to the leading contribution of the volume terms in the geometric picture, while the leading contribution to the surface terms is given by the standard cuscuton term $\sqrt{X}$. The trace of the extrinsic curvature $K_{ij}$ is denoted as $K=\gamma^{ij} K_{ij}$, with $\gamma_{ij}$ being the spatial metric, $\mathcal{R}$ is the intrinsic curvature (or the 3d Ricci scalar, $\gamma^{ij}{\cal R}_{ij}$), and $\mathcal{K}_{\text{GB}}$ is the boundary term for the 4-dimensional Gauss-Bonnet term \cite{Davis:2002gn}, given by
\begin{equation}\begin{split}
\mathcal{K}_{\text{GB}} = - 2 \qty(J -2 \varepsilon G^{ij} K_{ij} ),
\label{eq:cw3}
\end{split}\end{equation}
where
\begin{equation}\begin{split}
J=\frac{1}{3} \gamma^{ij}\qty(2 K K_{i c} K^c_{ \ j} + K_{c d} K^{cd} K_{ij} - 2 K_{i c} K^{cd} K_{dj}- K^2 K_{ij}).
\label{eq:cw3}
\end{split}\end{equation}
As illustrated in Fig. (\ref{fig:1}),
 in the discrete geometric picture, the effective couplings are constant coefficients that characterise the different phases or vacua separating each successive brane. 
 Each bulk contains distinct values for the cosmological constant, gravitational constant and couplings to the Lovelock terms. Likewise, every brane possesses unique values for the coefficients of the gravitational surface terms.  Whenever there is a discontinuity or jump in the bulk couplings, there will be a corresponding surface term associated with it. In the continuous limit, the couplings are taken to be slow-varying functions of the scalar field $\phi$. However, not all effective couplings are independent. The effective couplings corresponding to the boundary terms for the Einstein-Hilbert and Gauss-Bonnet terms are completely determined by the geometry \cite{Dyer:2008hb}. In the continuous limit, this fixes the coefficients $c_2, c_4, \cdots$:
\begin{align}
c_2(\phi) &= \lim_{\Delta\phi \to 0} 
 \frac{f(\phi+\Delta\phi)-f(\phi)}{\Delta\phi}=\frac{\pd f(\phi)}{\pd \phi},
 \\ 
 c_4(\phi) & = \lim_{\Delta\phi \to 0} 
 \frac{\omega(\phi+\Delta\phi)-\omega(\phi)}{\Delta\phi}=\frac{\pd \omega(\phi)}{\pd \phi}.
\label{eq:cpl}
\end{align}
In other words, the even-labelled surface couplings represent the rate of change of the bulk curvature couplings as we transition from one vacuum to another (in the limit where the sum becomes continuous). In this sense, the boundary is bookkeeping that something physical is changing.
We shall see that this will play a key role in ensuring, the ECT propagates only the two tensorial degrees of freedom. The reason is that in the continuous limit, these boundary contributions act as \textit{counterterms} to the bulk contributions, effectively eliminating the terms responsible for propagating the scalar degree of freedom. This will be clarified in the next section where we will closely examine the structure of the action in (\ref{eq:cw2a}). 

Note that the leading volume term, $V(\phi)$ describing the cuscuton potential,  has no derivatives and, therefore, no surface term associated with it. Finally, $c_3$ accompanies curvature terms that may live in each brane, which by dimensional analysis should be included in the effective action, such as $\mathcal{R}$.
In the limit of constant couplings, i.e., $ f=M_{Pl}^2$, where $M_{Pl} \equiv 1/\sqrt{8\pi G}$ is the reduced Planck mass and $c_1 = -\mu^2$, we recover the leading cuscuton action in (\ref{eq:cusc1}). It is now evident that the effective action in (\ref{eq:cw2a}) is simply an extension of (\ref{eq:cuscgem}) containing curvature corrections to the volume and the surface terms.

To conclude this section, we have constructed a geometric effective description for the cuscuton. Given that the original cuscuton theory (with only non-vanishing $V$ and $c_1$, and constant $f$) is a subset of the Type II MMG theories \cite{Lin:2017oow, Aoki:2018brq} or the covariant spatial gravity theories \cite{Gao:2019twq, Hu:2021yaq}, with only two tensorial degrees of freedom, it is natural to wonder whether ECT also fits into (either of) these two frameworks. We shall come back to this question in the next section, where we examine the ADM decomposition of ECT.

\section{ADM decomposition of action and EOMs}
\label{ADM}

As we saw in the previous section, the ECT is expressed as an expansion in the powers of curvature. While it would be desirable to confirm the validity of our conjecture for arbitrary powers of curvature,  the treatment of higher-order curvature invariants can become increasingly complex and technical. This would divert attention from the main focus of this manuscript which is to introduce the reader to the geometric construction of the ECT and outline its main features. While we shall briefly comment on the handling of higher curvature terms (such as $R_{GB}$) and some of the technicalities associated with the elimination of additional degrees of freedom beyond the two tensorial modes, we leave a more detailed analysis for future work. As such, for the remainder of this work, we will focus on the leading-terms in the ECT, namely
\begin{equation}\begin{split}
S_{{ECT}} & =  \Lambda^4\int \dd^4x \sqrt{-g}  \bigg[   f(\phi) \frac{^{(4)}R}{2 \Lambda^2} 
+ c_1(\phi) \frac{\sqrt{X}}{\Lambda} + V(\phi) +  \varepsilon c_2(\phi)  \frac{\sqrt{X} K}{\Lambda^2} \bigg].
\label{eq:ST}
\end{split}\end{equation}

\subsection{ADM decomposition}

The action in (\ref{eq:ST}) is given in terms of 3- and 4-dimensional objects. Ideally, it is desirable to express these terms using a consistent framework. To facilitate this, we opt to employ the ADM formalism \cite{Arnowitt:1962hi}.  The necessary ingredients of this decomposition are outlined below.

We define the normalisation $n^\alpha n_\alpha = \varepsilon = -1$ where $n^\alpha$ is a timelike vector pointing in the direction of increasing $\phi$, i.e., $n_\nu = \varepsilon \pd_\nu \phi /\sqrt{X}$. The induced metric is given by $g^{\alpha\beta} =  \gamma^{\alpha\beta} - n^\alpha n^\beta $ and we define the extrinsic curvature $K_{ab} =\nabla_a n_b - n_a a_b $ where $K_{ab} = K_{ba}$ is a symmetric tensor and acceleration $a_a= - n^b \nabla_b n^a= - D_a \ln{N} = - \frac{D_a N}{N}$. We will often make use of the identity $n^\mu \nabla_\alpha n_\mu = 0$
and that the extrinsic curvature satisfies the relation
 \begin{equation}\begin{split}
\mathcal{L}_{\vect{n}} \gamma_{ab} = 2 K_{ab}.
\label{eq:extrlie}
\end{split}\end{equation}
With these definitions, the projections of the 4-D Riemann tensor read (see also \cite{Deruelle:2009zk}, albeit with different conventions)
\begin{align}
    & ^{(4)} R_{ijkl} = K_{ik} K_{jl} - K_{il}K_{jk} + \, ^{(3)} R_{ijkl},
    \\
    &  ^{(4)} R_{ijk\vect{n}} \equiv n^\mu \, ^{(4)} R_{ijk\mu} = D_i K_{jk} - D_j K_{ik}, 
    \\
    & ^{(4)} R_{i\vect{n} k\vect{n}} \equiv n^\mu n^\nu \, ^{(4)} R_{i \mu j \nu} = -\mathcal{L}_\vect{n} K_{ij}  + K_{ik} K^{ \ k}_j + \frac{1}{N} D_i D_j N.
\end{align} 
Next, we use that $R_{\alpha\beta} = g^{\mu\nu} R_{\mu\alpha\nu\beta}$ and from this, we derive the projections of the 4-dimensional Ricci tensor:
\begin{align}
    & ^{(4)} R_{ij} = K K_{ij} - 2 K_i^{ \ a} K_{ja} + \, \mathcal{R}_{ij} - \frac{D_j D_i N}{N} + \mathcal{L}_{\vect{n}} K_{ij},
    \\
    &  ^{(4)} R_{\vect{n} j}  = D^a K_{ja} - D_j K, 
    \\
    & ^{(4)} R_{\vect{n} \vect{n}} = -K_{ij}K^{ij} + \frac{D_i D^i N}{N} - \mathcal{L}_{\vect{n}} K_{ij}, 
\end{align} 
Contracting again gives an expression for the Ricci scalar 
 \begin{equation}\begin{split}
^{(4)} R = K_{ij} K^{ij} + K^2 + \, \mathcal{R} - \frac{D^aD_a N}{N} + 2 \qty(\dot{K} -\mathcal{L}_{\vect{N}} K), 
\label{eq:rics}
\end{split}\end{equation}
where
 \begin{equation}\begin{split}
\mathcal{L}_\vect{N} K_{ij} \equiv N^k D_k K_{ij} + K_{ik} D_j N^k + K_{jk} D_i N^k, 
\label{eq:LN}
\end{split}\end{equation}
and $N^i$ is the shift vector. Finally, in the unitary gauge, i.e. $\phi = \phi(t)$, we see that under the ADM decomposition
\begin{equation}\begin{split}
\sqrt{X} = \mathcal{L}_{\vect{n}} \phi,
\label{eq:Xadm}
\end{split}\end{equation}
 where the Lie derivative acting in the direction of $\vect{n}$, is defined as 
\begin{equation}\begin{split}
\mathcal{L}_{\vect{n}} =\frac{1}{N} \tilde{\mathcal{L}}_{\vect{n}} =\frac{1}{N} \qty(\mathcal{L}_t - \mathcal{L}_{\vect{N}}),
\label{eq:ln}
\end{split}\end{equation}
with $\mathcal{L}_t \phi = \pd_t \phi$ and with our conventions\footnote{Here, we have assumed $\dot{\phi}>0$.}, we have that $\mathcal{L}_{\vect{N}}\phi = N^i \pd_i \phi  =  0$ so that we can safely ignore spatial derivatives acting on $\phi$.
Finally, in terms of the ADM variables, the extrinsic curvature and its trace can be written as follows: 
\begin{align}
   & K_{ij}  = \frac{1}{2N} \qty(\pd_t \gamma_{ij} - D_i N_j - D_j N_i),
   \\& K =  \gamma^{ij} K_{ij} = \frac{1}{2N} \qty(\gamma^{ij} \pd_t \gamma_{ij} - 2 D_i N^i ),
\end{align}
respectively. 

\subsection{ADM analysis}

Using the ADM definitions, given above, the action in (\ref{eq:ST}) can be expressed in the $3+1$
decomposition as
\begin{equation}\begin{split}
S_{{ECT}} & =  \Lambda^4\int \dd^4x \, N\sqrt{\gamma}  \bigg[ V(\phi)  + c_1(\phi) \frac{\sqrt{X}}{\Lambda} +  \frac{f(\phi)}{2 \Lambda^2} \Big(K_{ab} K^{ab} + K^2 + \,\mathcal{R} 
\\&- \frac{2 D^{a}D_{a} N}{N} 
 + 2 \mathcal{L}_\vect{n} K \Big) -  \varepsilon c_2(\phi)  \frac{\sqrt{X} K}{\Lambda^2} \bigg].
\label{eq:sadm}
\end{split}\end{equation}
The term proportional to $D^a D_a N$ is independent of the volume form and therefore  a total derivative, while the term proportional to $\mathcal{L}_\vect{n} K$ can be integrated by parts to give 
\begin{equation}\begin{split}
S_{{ECT}} & =  \Lambda^4\int \dd^4x \, N\sqrt{\gamma}  \bigg[ V(\phi)  + c_1(\phi) \frac{\sqrt{X}}{\Lambda} +  \frac{f(\phi)}{2 \Lambda^2} \Big( K^2 -K_{ab} K^{ab} + \,\mathcal{R} \Big) 
\\& - \frac{1}{\Lambda^2} f'(\phi) \sqrt{X}  K
-  \varepsilon c_2(\phi)  \frac{\sqrt{X} K}{\Lambda^2} \bigg].
\label{eq:sadm1}
\end{split}\end{equation}
Then,  by using (\ref{eq:cpl}), (\ref{eq:Xadm}) and that with our conventions $\varepsilon = -1$, it is straightforward to show that the last two terms cancel out, leaving
\begin{equation}\begin{split}
S_{{ECT}} & =  \Lambda^4\int \dd^4x \, N\sqrt{\gamma}  \bigg[ V(\phi)  + c_1(\phi) \frac{\sqrt{X}}{\Lambda} +  \frac{f(\phi)}{2 \Lambda^2} \Big( K^2 -K_{ab} K^{ab} + \,\mathcal{R} \Big) \bigg].
\label{eq:sadm2}
\end{split}\end{equation}
Given that $\phi=\phi(t)$ and $\sqrt{X} = \dot{\phi}(t)/N$ in the unitary gauge, the action (\ref{eq:ST}) written in the form (\ref{eq:sadm2}) is nothing but a subset of the minimally-modified gravity theories MMG-II \cite{DeFelice:2020eju, Lin:2017oow}  having the generic form $L= N F(K_{ij}, R_{ij}, \gamma^{ij}, t) + G(K_{ij}, R_{ij}, \gamma^{ij}, t)$.  Comparing the two, we see that the potential and spatial curvature terms, which are coupled to $f(\phi)$, appear to be linear in the lapse $N$ and therefore correspond to the term $N F$, while the cuscuton term, which couples  to $c_1(\phi)$,
 is independent of the lapse and therefore corresponds to the term $G$.   The classification II refers to the fact that these type of modified gravity theories have no Einstein frame. In \cite{DeFelice:2020eju, Lin:2017oow} it was shown, through Hamiltonian analyses, that this type of action propagates only two tensorial degrees of freedom\footnote{See Sections 3.1 and 3.4 in \cite{Lin:2017oow}, for a proof of this statement }.

We can use the geometric construction of the ECT to obtain some intuition of why this is the case. Going from (\ref{eq:sadm1}) to (\ref{eq:sadm2}) we found that the contributions from the surface terms, in the geometric picture, act as \textit{counterterms}, i.e., they induce cancellations such that the resulting system does not propagate a scalar degree of freedom, but simply leads to a set of constraint equations. 
This becomes clearer when examining the equations of motion:
\begin{align}
\label{hami}
    0&= V(\phi)  - \frac{f(\phi)}{2} \qty( K^2 - K_{cd}K^{cd} -\,\mathcal{R} ), 
    \\
    \label{mom}
    0 & = f(\phi) \qty(D_c K_i^{ \ c} - D_i K),
    \\
    \label{evol}
    0 & =   V(\phi) \gamma_{ij} + c_1(\phi) \sqrt{X} \gamma_{ij} + f(\phi) \bigg[K K_{ij} - 2 K_{i}{}^{a} K_{ja} - \, G_{ij}  \nonumber
\\& - \frac{1}{2} K_{ab} K^{ab} \gamma_{ij} -\frac{1}{2} K^2 \gamma_{ij}  
- \gamma_{ij} \mathcal{L}_{\vect{n}} K + \mathcal{L}_{\vect{n}} K_{ij} \bigg] \nonumber
\\& +  f'(\phi) \mathcal{L}_\vect{n} \phi \qty(K_{ij} - K \gamma_{ij}), 
\\
\label{phiEOM}
0 & =   V' (\phi)  - c_1(\phi) K \frac{\mathcal{L}_{\vect{n}} \phi}{\sqrt{X}}  + \frac{f'(\phi)}{2 } \Big( K^2 -K_{ab} K^{ab} + \,\mathcal{R} \Big),
\end{align}
where (\ref{hami}) is the Hamiltonian constraint, (\ref{mom}) is the momentum constraint, (\ref{evol}) is the evolution of the spatial metric and (\ref{phiEOM}) is the scalar field equation. For brevity, we have absorbed the scale $\Lambda$ within the couplings. 
Equation (\ref{evol}) is not independent; therefore, we can directly substitute the Hamiltonian constraint (\ref{hami}) into the scalar field equation (\ref{phiEOM}), to find a constraint equation:
\begin{equation}\begin{split}
K = \frac{1}{c_1(\phi)} \qty[\frac{3}{2} \qty( V'(\phi) + f'(\phi) \mathcal{R}) ].
\label{eq:Kconstr}
\end{split}\end{equation}
For a flat FLRW background, the equations of motion read
\begin{align}
\label{fried1bg}
   0 & = 3 H^2 f(\phi) - V(\phi),
   \\
   \label{fried2bg}
  0 & = f(\phi) \qty(\dot{H} + \frac{3}{2} H^2) + H f'(\phi) \dot\phi - \frac{1}{2} c_1(\phi) \dot\phi - \frac{1}{2} V(\phi),
  \\
  \label{cuscbg}
0&= 3 H^2 f'(\phi) - 3  \, c_1(\phi) \, H  + V'(\phi).
\end{align}
If we add matter into the mix, we find that the second Friedmann equation is indeed independent. As before, we can substitute the first Friedmann equation into the scalar field equation to find a constraint equation: 
\begin{equation}\begin{split}
V'(\phi) - c_1(\phi) \sqrt{\frac{3 \, V'(\phi)}{4 \, f'(\phi)}} =0.
\label{eq:STconstrBG}
\end{split}\end{equation}
Similarly to the case of cuscuton, the ECT field equation in  empty flat FLRW reduces to a constraint on the functions $c_1(\phi), f(\phi) $, and $V(\phi)$, without actually constraining the expansion history. Therefore, we only expect a sensible cosmology, if we introduce additional matter that changes the left-hand sides of equations (\ref{fried1bg}-\ref{fried2bg}) . In this case, the field $\phi$ will simply follow the expansion rate $H$, through the field equation (\ref{cuscbg}).

\section{Going beyond leading-order in the effective cuscuton theory}
\label{beyond}

As we discussed earlier, the ECT is an expansion in powers of curvature, allowing one to incorporate an arbitrary number of higher-derivative  terms into the action, accompanied by the corresponding counterterms, as prescribed by the geometric picture. Supposing that our conjecture proves valid,  these counterterms should eliminate the propagation of the scalar degree of freedom {\it at every order}, thereby ensuring that the theory retains solely the two desired tensorial degrees of freedom. As with all EFTs, in practise, the theory is truncated at a specific order within the perturbative expansion. This is typically decided depending on two factors: i) the nature of the intended physical application, and ii) the energy range that our physical experiments can feasibly probe \cite{burgess2020introduction}.

When extending our EFT beyond the leading order, one possible option would involve including curvature invariants, such as those discussed in \cite{Weinberg:2008hq, Solomon:2017nlh}. Unfortunately, not all boundary terms for curvature invariants have yet been fully explored \footnote{It was recently suggested that they make take the form as in \cite{Neri:2023esr}.}. 
For this reason, here we choose to work with terms of Lovelock type, such as the Gauss-Bonnet term whose boundary contributions to the action were previously studied in \cite{Davis:2002gn}. As an added bonus, this choice guarantees second-order equations of motion without the need of performing field redefinitions, order by order in the preturbative expansion, which is typically the case with other curvature invariants. 

Including the Gauss-Bonnet (ignoring the $c_3$ term for now), the action (\ref{eq:cw2a}) becomes
\begin{equation}\begin{split}
S_{{ECT}} & =  \Lambda^4\int \dd^4x \, N\sqrt{\gamma}  \bigg[ V(\phi)  + c_1(\phi) \frac{\sqrt{X}}{\Lambda} +  \frac{f(\phi)}{2 \Lambda^2} \Big( K^2 -K_{ab} K^{ab} + \,\mathcal{R} \Big)
\\& + \frac{\omega(\phi) }{ 2\Lambda^4} \Big( \mathcal{R}_{GB} -12 K_{a}{}^{c} K^{ab} K_{b}{}^{d} K_{cd} - \frac{8}{3} K K_a^{ \ c} K^{ab} K_{cd} - K_{ab} K^{ab} K_{cd} K^{cd} 
 \\& +2 K^2 K_{cd} K^{cd} - \frac{1}{3} K^4 
 + 16 K_{a}{}^{c} K^{ab} \,\mathcal{R}_{bc}  
  + 2 K_{ab} K^{ab} \,\mathcal{R} - 2 K^2 \,\mathcal{R} 
    \\& +  4 K^{ab} K^{cd} \,\mathcal{R}_{acbd} 
  + 8 \, D_{b}D_{a} G^{ab}  + 8 K^{ab} D_b D_a K + 8 K D_c D_b K^{bc}  -8 K D_{c}D^{c} K
 \\&     - 16 K^{ab} D_{(c}D_{b)} K_{a}{}^{c} + 8  K^{ab} D_{c}D^{c}K_{ab}  
  + 8 K_{ab} \mathcal{L}_{\vect{n}} G^{ab}  \Big) \bigg],
\label{eq:sadmGBibp}
\end{split}\end{equation}
where we have performed several integrations by parts and used the relation in (\ref{eq:cpl}) to cancel out the counterterms. We have defined the following:
\begin{align}
\label{eq:3GB}
   \mathcal{R}_{GB} & = \mathcal{R}^2 - 4 \mathcal{R}^{ij} \mathcal{R}_{ij} + \mathcal{R}^{ijkl} \mathcal{R}_{ijkl}, 
   \\
   \label{eq:3Gij}
   G_{ij} & = R_{ij} - \frac{1}{2} \gamma_{ij} R,
\end{align}

and
\begin{equation}\begin{split}
A_{(ab)} \equiv \frac{1}{2} \qty(A_{ab}+ A_{ba}),
\label{eq:ident}
\end{split}\end{equation}
is defined as the symmetrized sum for any two tensor indices. 

Comparing this to the action in (\ref{eq:sadm2}) and employing the language used in \cite{Lin:2017oow, Gao:2019twq}, the action in (\ref{eq:sadmGBibp}) is of the general form
%
\begin{equation}\begin{split}
S_{ECT} = \Lambda^4\int \dd^4x \, N\sqrt{\gamma}   F(K_{ij}, R_{ij}, D_i, \gamma^{ij}, \mathcal{L}_\vect{n}, t).
\label{eq:GBact}
\end{split}\end{equation}
We anticipate, the presence of mixed space-time derivatives, (i.e., of the form $D_i K$) will likely modify the Hamiltonian structure of (\ref{eq:sadmGBibp}) and make it challenging to solve conditions obtained from Hamiltonian analysis, as discussed in \cite{Lin:2017oow, Gao:2019twq}.  Certainly, we did not face these issues with the action in (\ref{eq:sadm2}). 
Additionally, the presence of the Lie derivative in the last term of (\ref{eq:sadmGBibp}), namely $\mathcal{L}_\vect{n} G^{ab}$, introduces complications and  a Hamiltonian analysis may not be feasible due to the mixing of time and spatial partial derivatives of the metric. Fortunately for us, the Lie derivative drops out at the level of the equations of motion. This can be seen by following the same procedure as in the previous section. We can extend the constraint equation in (\ref{eq:Kconstr}) to include the contributions from the Gauss-Bonnet term plus counterterms, which reads
\begin{align}
    K = \frac{1}{c_1(\phi)} \qty[ \frac{3}{2} V'(\phi) + f'(\phi) \mathcal{R} + \omega'(\phi) (  B + C ) ],
\end{align}
where
\begin{align}
  B &=   12 K_{a}{}^{c} K^{ab} K_{b}{}^{d} K_{cd} + \frac{8}{3} K K_a^{ \ c} K^{ab} K_{cd} + K_{ab} K^{ab} K_{cd} K^{cd} 
  -2 K^2 K_{cd} K^{cd} +\frac{1}{3} K^4 ,
 \\
 C & = \mathcal{R}_{GB}   + 8 \, D_{b}D_{a} G^{ab}+ 16 \frac{D_c D_a N}{N} \Big(K_a^{ \ c} K^{ab} - K K^{bc} \Big) 
    + 8 \frac{D_c D^c N}{N} \Big( K^2 - K_{ab} K^{ab} \Big). 
\end{align}
Indeed, we discover that the troublesome Lie derivatives can be eliminated from the constraint equation.  The terms involving spatial covariant derivatives acting on the lapse were initially removed from the action in (\ref{eq:sadmGBibp}), through repeated integration by parts.  However, these terms have been reintroduced into the equations of motion, as one would expect. Nonetheless, the eliminations of $\mathcal{L}_\vect{n} G^{ab}$  from the constraint equation, gives us hope that a proper Hamiltonian analysis may be possible\footnote{We thank Shinji Mukohyama for bringing this point to our attention.}.

\section{Discussion}

\label{discus}

In the Introduction, we discussed various approaches to constructing cuscuton-like theories, which have been demonstrated to be helpful in obtaining theories of modified gravity without an additional scalar degree of freedom. These could have implications in understanding dark energy, black holes, or even the cosmological Big Bang. Yet, it is crucial to maintain focus on the fundamental aspects defining what makes a theory cuscuton-like. This is particularly intriguing since 
the cuscuton appears in several places in physics, such as being the IR limit to Horava-Lifshitz gravity and  the UV limit ($X \gg 1$) of the (anti-)DBI action. 

Understanding the fundamental reasons behind a theory being cuscuton-like might hold the key to discovering something much more profound. This is the motivation of this work, in which we extended the cuscuton theory through geometric postulates by introducing curvature corrections to both the volume and the surface terms in the original cuscuton action. Our assumptions involve a stack of spacelike branes, separated by 4-dimensional bulks, conjecturing that the cuscuton, initially a discrete field, becomes continuous in the limit; there are many such transitions.

Analysis of this effective cuscuton theory (ECT), at the quadratic level of the action, reveals that it is a subset of Type II MMG or covariant spatial gravity theories. These are modified gravity theories that propagate only the two tensorial degrees of freedom.  We find that, in the continuous limit, the surface contributions act as ``counterterms'', inducing cancelations that effectively eliminate terms responsible for propagating the scalar degree of freedom.

This novel approach to constructing cuscuton-like theories serves as the manuscript's focus, introducing readers to its main mechanics and features. Detailed treatments of higher-order curvature terms, such as the Gauss-Bonnet, are deferred to future work, along with the exploration of underlying symmetries and the stability analysis of this geometric construction.

Finally, the geometric picture for the cuscuton involves a series of subsequent phase transitions (or quantum tunnelings), which are commonly believed to become quickly non-perturbative through bubble nucleations and collisions. One may, however, imagine that there may be an alternative regime in this process that may be ``cuscuton-like", but we shall refrain from commenting any further at this point, as we enter the realm of pure speculation.

\subsection*{Acknowledgements} 

The authors would like to thank C.P. Burgess, X. Gao, J-O. Gong, W. Kinney, T. Kobayashi, S. Mukohyama, M. Sasaki and G. Tasinato   for helpful discussions and insightful comments to the manuscript. NA is supported by the University of Waterloo, Natural Sciences and Engineering Research
Council of Canada (NSERC) and the Perimeter Institute for Theoretical Physics. Research at
Perimeter Institute is supported in part by the Government of Canada through the Department
of Innovation, Science and Economic Development Canada and by the Province of Ontario
through the Ministry of Colleges and Universities. MM is  supported in parts by the Mid-Career Research Program (2019R1A2C2085023) through the National Research Foundation of Korea Research Grants. MM is also supported Kavli IPMU  which was established by  the World Premier International Research Center Initiative (WPI), MEXT, Japan.

	\bibliographystyle{JHEP}
	\bibliography{bib.bib}

\end{document}